\documentclass[a4,twocolumn,showpacs,showkeys]{revtex4}
\usepackage{graphicx}
\usepackage{amsmath}
\usepackage{amssymb}
\usepackage[utf8x]{inputenc}

\def\be{\begin{equation}}
\def\ba{\begin{eqnarray}}
\def\ee#1{\label{#1}\end{equation}}
\def\ea#1{\label{#1}\end{eqnarray}}
\def\la{\langle}
\def\ra{\rangle}
\def\bs{\begin{center}}
\def\es{\end{center}}

\begin{document}

\title{Negative conductances of  Josephson junctions:  Voltage fluctuations and energetics} 

\author{L. Machura}
\affiliation{Institute of Physics, University of Silesia,
40-007 Katowice, Poland}

\author{M.  Kostur}
\affiliation{Institute of Physics, University of Silesia,
40-007 Katowice, Poland}

\author{P. Talkner}
\affiliation{Institute of Physics, University of Augsburg,
D-86135 Augsburg, Germany}
\author{P. H\"anggi}
\affiliation{Institute of Physics, University of Augsburg,
D-86135 Augsburg, Germany}

\author{J. \L uczka\footnote{Corresponding author. E-mail: jerzy.luczka@us.edu.pl }}
\affiliation{Institute of Physics,  University of Silesia,
40-007 Katowice, Poland}

\begin{abstract}
We study a resistively and capacitively 
shunted Josephson junction,
which is driven by a combination of time-periodic and constant
currents.  Our investigations concern three main problems: (A) The voltage
fluctuations across the junction;  (B) The quality of transport expressed in terms of the P\'eclet number;   (C) The efficiency of energy transduction  
from external  currents. These issues are discussed in different
parameter regimes that lead to: (i) absolute negative
conductance; (ii) negative differential
conductance, and (iii) normal, Ohmic-like conductance.  Conditions for optimal operation 
of the system are studied. 

\end{abstract}

\keywords {Josephson junction, anomalous transport, negative conductance, diffusion, voltage fluctuations,   efficiency}  

\pacs{
74.25.Fy 
85.25.Cp 
73.50.Td    
05.45.-a, 
}

\maketitle


\section{Introduction}

Transport processes in periodic systems play an important role 
in   a great variety of everyday life phenomena.     
Two prominent examples are the  electric transport in metals providing
a prerequisite of  modern civilization, and the movement 
of so-called molecular motors 
(like kinesin and dynein) along microtubules in biological cells which
are of crucial relevance for the functioning of any higher living organism.
Josephson junctions belong to the same class of systems being
characterized by a spatially periodic structure.  In the limiting case
of small tunnel contacts the mathematical description of a Josephson
junction is identical to that of a Brownian particle moving in a
periodic potential. Such models have also frequently been
employed under nonequilibrium conditions to
describe Brownian ratchets and molecular motors,
see Refs  \cite{hanRev,motor} and references therein.  
Of particular importance for technological applications are 'rocked' thermal Brownian motors
operating either in  overdamped or  underdamped regimes, e.g. see in Ref.  \cite{rocked}. 
The majority of papers on transport in periodic systems are focused on the asymptotic
long time behavior of averaged quantities such as the mean velocity
of a molecular motor, or the mean voltage drop in a Josephson contact
\cite{ratchet}. The main emphasis of these works lies  on formulating
and exploring    conditions that are necessary for the generation and control of
transport, its direction, 
and magnitude as well as its dependence on system parameters like temperature
and external load. Apart from these well investigated questions  
other important features concerning the {\it   quality} of transport though have remained 
unanswered to a large extent.  The key to
these problems lies in the investigation of the {\it fluctuations} about the
average asymptotic behavior \cite{machuraJPC}. 

In the present paper we
continue our previous studies on anomalous electric transport in driven,
resistively and capacitively shunted Josephson junction devices  
\cite{machuraPRL,aip,kosturPRB}. 
These  investigations were focused on the current-voltage
characteristics, in particular on  negative conductances.  In contrast, in the present paper we
investigate the fluctuations of voltage, the  diffusion processes of
the Cooper pair phase difference across a Josephson junction as well
as the energetic performance of such a device. 

The paper is organized as follows. In the next 
section,  we briefly describe  the Stewart-McCumber model for the dynamics
of the voltage  across a junction. In Section 3, we study voltage fluctuations, 
phase difference diffusion,  and the efficiency of the device.
Conclusions are contained in  Section 4.

\section{Model of  resistively and capacitively shunted Josephson
  junction}

The Stewart-McCumber model describes the
semi-classical regime of a small (but not ultra small) Josephson
junction for which a spatial dependence of characteristics can be
neglected. The model contains three additive current contributions:
a Cooper pair tunnel current characterized by the critical current
$I_0$, a normal (Ohmic) current characterized by the normal state
resistance $R$ and a displacement current due to the capacitance $C$ of
the junction. Thermal fluctuations of the current are taken into
account according to the fluctuation-dissipation theorem and satisfy the Nyquist
formula associated with the resistance $R$. The quasi-classical
dynamics of the phase difference $\phi=\phi(t)$ between the
macroscopic wave functions of the Cooper pairs on both sides of
the junction is described by the following equation
\cite{barone,kautz},
\begin{eqnarray} \label{JJ1}
\frac{\hbar}{2e}  C\:\ddot{\phi} +  \frac{\hbar}{2e}  \frac{1}{R} \dot{\phi}
+  I_0 \sin (\phi) =  I_d  \nonumber\\
+ I_a \cos(\Omega t+\varphi_0) + \xi (t)\;, 
\end{eqnarray}
where the dot denotes the differentiation with respect to time, $I_d$ and
$I_a$ are the amplitudes of the applied direct (dc) and alternating (ac) 
currents,
respectively, $\Omega$ is the angular frequency and $\varphi_0$
defines the initial phase value of the ac-driving.  
Thermal equilibrium fluctuations are modeled by zero-mean 
Gaussian white noise $\xi(t)$ with   the correlation function  
$<\xi(t) \xi(s)> = (2 k_B T/R) \:\delta(t-s)$,  where  
$k_B$ is the Boltzmann constant and $T$ is  temperature of the
system. 

The limitations of the Stewart-McCumber model and its range of
validity are discussed e.g. in Sec. 2.5 and 2.6 of Ref.  \cite{kautz}.
There are various other physical systems that are described by
Eq. (\ref{JJ1}).  A typical example is a Brownian particle moving in the
spatially periodic potential $U(x)=U(x+L)= - \cos(x)$ of period $L=
2\pi$, driven by a time-periodic force and a constant force
\cite{machuraJPC}.  In this case, the variable $\phi$ corresponds to the spatial
coordinate $x$ of the Brownian particle and ac and dc
play the role of periodic driving and a static tilt force,
respectively, acting on the particle. Other specific
systems are: a pendulum with an applied torque \cite{barone}, rotating
dipoles in external fields \cite{Reg2000,Coffey}, superionic
conductors \cite{Ful1975} and charge density waves \cite{Gru1981}.

It is convenient to transform Eq. (\ref{JJ1}) to a dimensionless
form.  We rescale the time $t'=\omega_p t$, where
$\omega_p=(1/\hbar)\sqrt{8 E_JE_C}$ is the Josephson plasma frequency
expressed by the Josephson coupling energy $E_J=(\hbar/2e)I_0$ and the
charging energy $E_C=e^2/2 C$. Then Eq. (\ref{JJ1}) takes the form
\cite{barone,kautz}
\begin{eqnarray}
\label{JJ2} \frac{d^2 \phi}{dt'^2} + {\gamma} \frac {d  \phi}{d t'}
+ \sin (\phi)  = i_0 
+ i_1  \cos(\Omega_1 t' +\varphi_0)
 \nonumber \\+ \sqrt{2 \gamma D} \; \Gamma(t').
\end{eqnarray}
The dimensionless damping  constant ${\gamma} = 1/\omega_p RC$ is
given by the ratio of two characteristic times: $\tau_0=1/\omega_p$
and the relaxation time $\tau_r = RC$. This damping  constant 
$\gamma$ measures the strength of dissipation. The ac amplitude and 
angular frequency are $i_1 = I_a / I_0$ and
$\Omega_1 = \Omega \tau_0=\Omega/\omega_p$, respectively.  The
rescaled dc strength reads $i_0= I_d/ I_0$. The rescaled zero-mean
Gaussian white noise $\Gamma(t')$ possesses the auto-correlation
function $\langle \Gamma(t')\Gamma(u)\rangle=\delta(t'-u)$, and the
noise intensity $D = k_B T / E_J$ is given as the ratio of two
energies, the thermal energy and the Josephson coupling energy
(corresponding to the barrier height).

Because Eq.  (\ref{JJ2}) is equivalent to a set of three autonomous
first order ordinary differential equations, the phase space of (\ref{JJ2}) is
three-dimensional. For vanishing diffusion constant, $D=0$, the system
becomes deterministic.  The resulting deterministic nonlinear dynamics ($D = 0$)
exhibits a very rich behavior ranging from periodic to quasi-periodic
and chaotic solutions in the asymptotic long time limit. Moreover,
there are regions in parameter space where several attractors coexist.
In the presence of small noise these attractors still dominate the
dynamics in the sense that most of the time the trajectory stays close
to one of these attractors. Only rarely, transitions between the
attractors take place. So, the locally stable states of the noiseless
dynamics become metastable states in the presence of weak noise. 
Apart from that, the presence of noise may also let the system come 
close to deterministic unstable orbits which it may follow for quite some time. 

Strictly speaking, the deterministic regime $D=0$ is only reached in
the limit of zero temperature for which quantum effects become
relevant. These are not taken into account in the classical Langevin
equation (\ref{JJ2}). However, for sufficiently large Josephson
junctions  a region of low temperatures exists
for which both thermal and classical fluctuations can be neglected on
those time scales that are experimentally relevant.

The averaged transport behavior is completely determined by the
current--voltage characteristic, i.e. the functional dependence of the
averaged voltage on the applied dc--strength in the asymptotic limit of
large times when all transient phenomena have died out.   
To obtain this current--voltage characteristic, we numerically simulated $10^3$  solutions of Eq.
(\ref{JJ2}) from which we  estimated   the stationary dimensionless voltage defined as
\begin{equation} \label{v}
v=\langle \dot \phi(t') \rangle,
\end{equation}
where the brackets denote averages (i) over the initial conditions $(
\phi(0), \dot\phi(0), \varphi_0)$ according to a uniform distribution
on the cube $\{\phi(0)\in [0,2 \pi], \dot{\phi}(0)\in [-2,2], \varphi_{0}
\in [0,2 \pi] \}$, (ii) over realizations of thermal noise $\Gamma(t')$
and (iii) a temporal average over one cycle
period of the external ac-driving once  the result of the first two
averages  have evolved  into a
periodic function of time. The stationary physical voltage is then expressed as
\begin{equation}
\label{V}
V= \frac{\hbar \omega_p}{2e} \; v.
\end{equation}
For a vanishing dc--strength, $i_{0}=0$, also the average voltage must vanish
because under this condition Eq. (2) as well as the probability
distribution with respect to which the average is performed are  
invariant under the transformation 
$(\phi, \varphi_0) \to (-\phi, \varphi_0 + \pi)$. For  non-zero  currents $i_0
\ne 0$, this symmetry is broken and the
averaged voltage can take non-zero values, which typically assume the
same sign as the bias current $i_0$. 
Apart from this ``standard'' behavior, a Josephson junction
may also exhibit other more exotic features, such as absolute negative
conductance (ANC) \cite{machuraPRL,aip}, negative differential
conductance (NDC), negative--valued nonlinear conductance (NNC) and 
reentrant effects into states of negative conductance
\cite{aip,kosturPRB}.  In mechanical, particle-like motion terms,
these exotic transport 
patterns correspond to different forms of negative
mobility of a Brownian particle.

\begin{figure}
\includegraphics[scale=0.6]{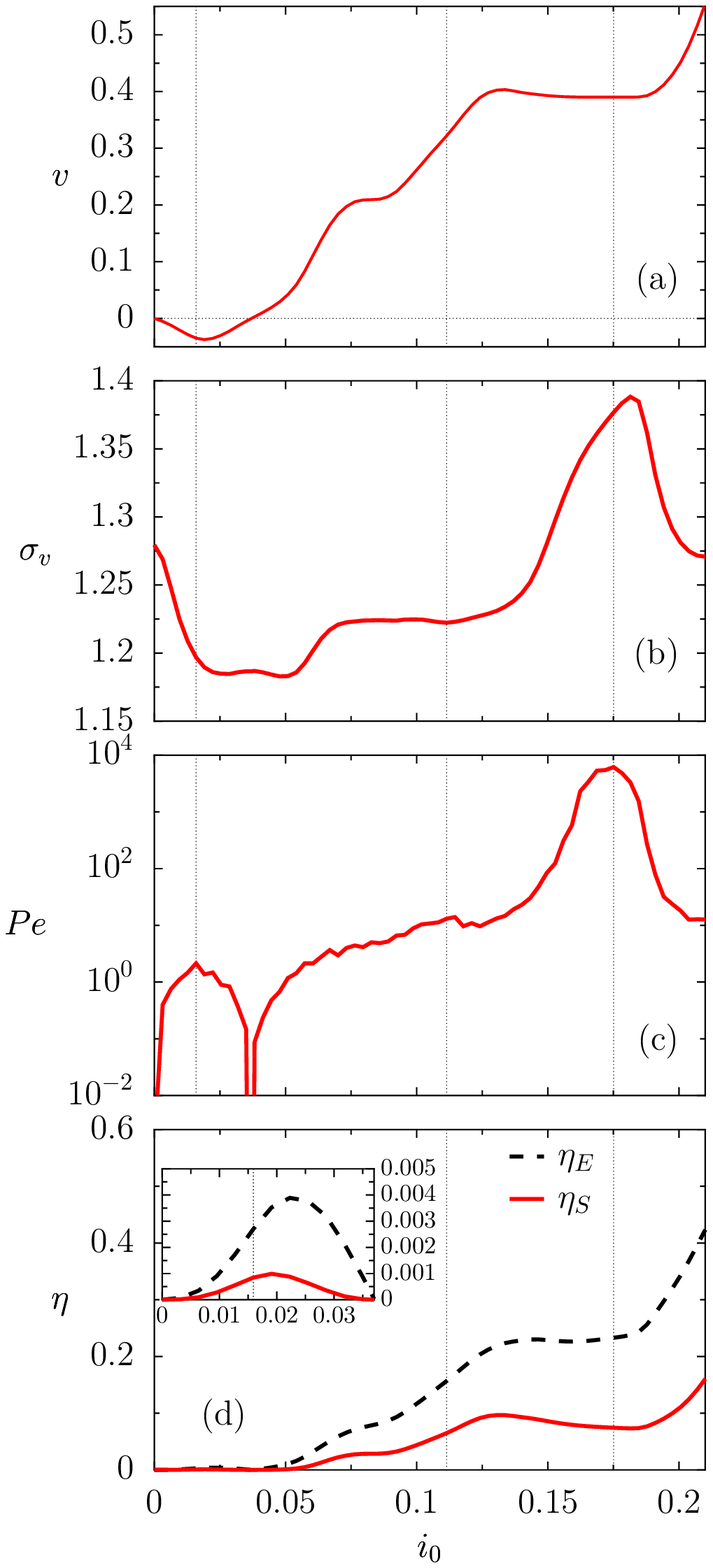} 
\caption{(color online) Various performance measures of a
  Josephson  junction in dependence of the dc--strength are compared.
In panel (a) the stationary voltage $v$ defined in Eq. (\ref{v}) is
depicted while panels (b), (c) and (d) display the standard deviation
$\sigma_v$ as a measure the voltage fluctuations, cf. eq. (\ref{sigma}),
the P\'eclet number  Pe  (Eq. (\ref{factors})) 
and the two efficiency measures
$\eta_{E}$ (Eq. \ref{EtaEnergy}) and $\eta_{S}$  (Eq. \ref{EtaStokes}), respectively. 
The remaining system's  parameters are: the thermal noise intensity $D=0.001$, ac driving
  frequency $\Omega=0.78$, ac driving amplitude $i_1=0.732$ and
  damping constant $\gamma=0.143$.  Three thin vertical lines 
  mark the three dc--strengths 
  $i_0=0.0159$ (regime of absolute negative conductance), $i_0=0.1114$
  (Ohmic-like regime, chaotic regime, large diffusion, relatively
  small current fluctuations), $i_0=0.175$ (regime of negative
  differential conductance, very regular motion, small diffusion but
  relatively large current fluctuations). For corresponding trajectories
  and phase portraits see Figs. \ref{fig2}, \ref{fig3} and
  \ref{fig4}.  }
\label{fig1}
\end{figure}

\section{Transport characteristics} 

Apart from the averaged stationary velocity $v$, which presents the basic 
transport measure, there are other quantities that characterize the random
deviations of the voltage about its average $v$ at large times such as
the voltage variance 
\be
\sigma_v^2 = \la \dot\phi^2 \ra -\la \dot\phi \ra^2.
\ee{sigma}
Here
the average is performed with respect to the same probability
distribution as for $v$ in Eq. (\ref{v}).
This variance determines the range  
\be
v(t') \in \left( v  -\sigma_v,  v+\sigma_v\right)
\ee{v(t)}
of the 
dimensionless voltage $v(t')=\dot{\phi}(t')$  in which its actual value is
typically found.
Therefore the voltage may assume the opposite
sign to  the average voltage $v$ if $\sigma_v > v $.

In order to quantify the effectiveness of a device in terms of the
power output at a given input, several measures have been proposed in
the literature \cite{bier,Sekimoto2000,Suzuki2003,wang,machuraPRE,rozen}.
Here we discuss 
two of them, which yield consistent results. 
 For the systems described by Eq. (\ref{JJ2}), the  {\it efficiency of
 energy conversion} is defined as the  ratio of the power $P=i_0 v$
 done against an external  bias $i_0$ and 
the input power $P_{in}$  \cite{sintes,physA},
\be \eta_{E} = \frac{|i_0 v|}{P_{in}},   \ee{EtaEnergy}
where $P_{in}$ is the total ac and dc power supplied to the system.
It is given by \cite{machuraPRE}
\be
 P_{in}   = \gamma [v^2  + \sigma_v^2 - D]  = \gamma [\la \dot\phi^2 \ra - D]. 
\ee{eta}
This relation follows from an energy balance of the underlying
 equation of motion  (\ref{JJ2}). 
Further, the  {\it Stokes efficiency}  is given by the relation \cite{wang} 
\be 
\eta_{S} =  \frac{i_{\gamma} v}{P_{in}}=\frac{\gamma v^2}{P_{in}} 
= \frac{ v^2}{v^2  + \sigma_v^2 - D},   
\ee{EtaStokes}
where  $i_{\gamma} = \gamma v$ denotes the Ohmic current, cf. Eq. (\ref{JJ2}). 
In contrast to the definition of the efficiency of energy conversion $\eta_E$, the definition 
of the Stokes efficiency $\eta_S$  does not  contain  the damping  constant $\gamma$. 
We note that a decrease of the voltage variance $\sigma_v^{2}$ leads
to a smaller input power and hence to an increase of the energetic
efficiency. 

Another quantity that characterizes the effectiveness of transport is
the effective diffusion coefficient of the phase difference $\phi(t)$,
describing the spreading of 
trajectories and fluctuations around the average phases. 
It is defined as follows
\be D_{\phi} = \lim_{t \rightarrow \infty} \frac{\la \phi^2(t) \ra - \la
\phi(t) \ra^2}{2t}.  \ee{Deff}
The coefficient $D_{\phi}$ can also be introduced via a generalized
Green-Kubo relation \cite{machuraJPC}.  Intuitively, the
diffusion coefficient is small and the transport is more effective if
the stationary 
voltage is large and the spread of trajectories is small.
The ratio $D_{\phi}/2\pi$ can be considered as a velocity characterizing the
phase difference diffusion over one period. Its relation to the
averaged velocity $v$ of the phase difference  determines the
dimensionless P\'eclet number Pe defined as
\ba \mbox{Pe} = \frac{ 2\pi|\la v \ra| }{D_{\phi}}. 
 \ea{factors}
A large P\'eclet number indicates a motion of mainly regular
nature.  If it is small then random or chaotic influences dominate the dynamics.      

\begin{figure}
\includegraphics[width=1\linewidth]{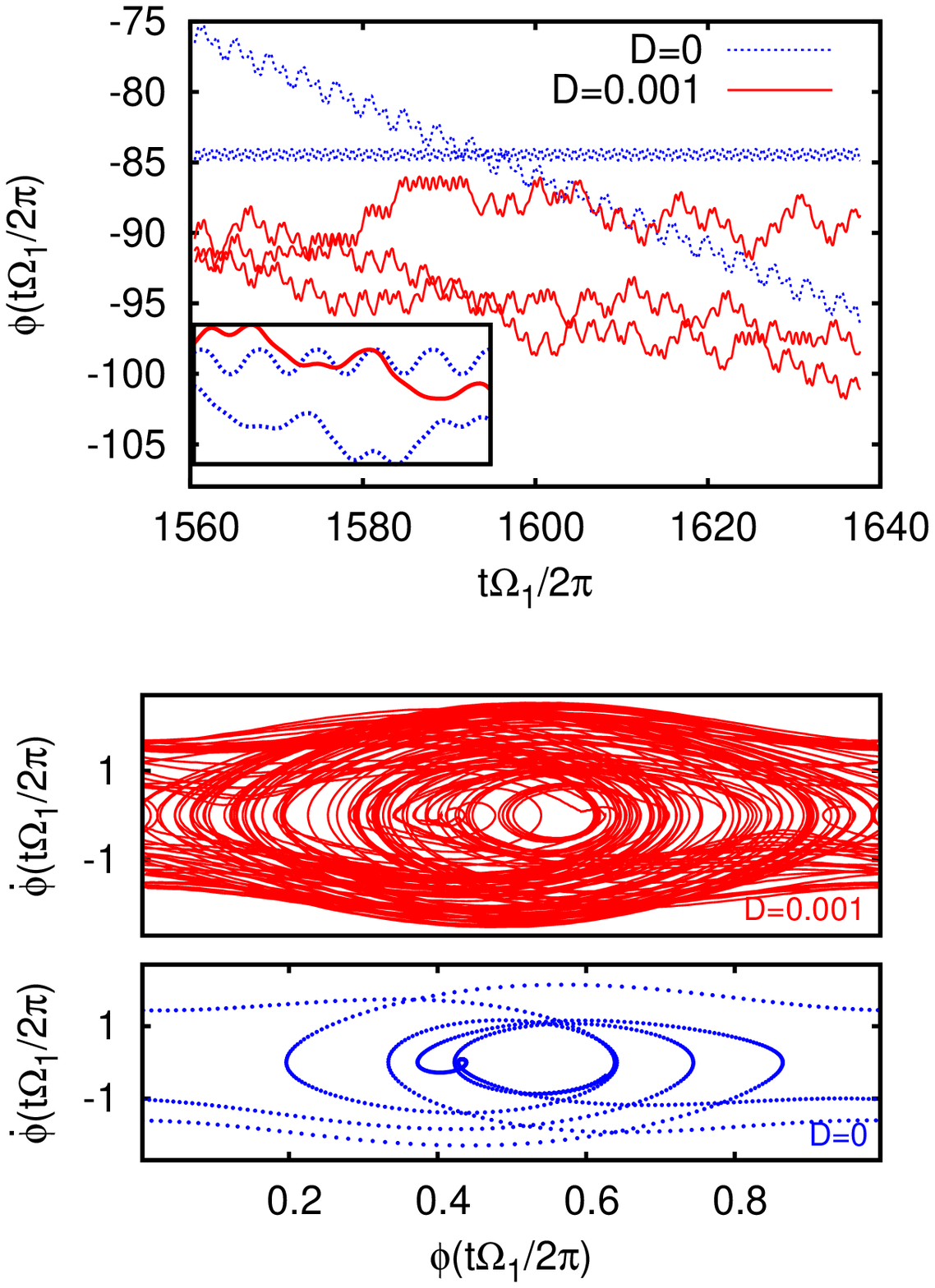}
\caption{(color online) 
In the ANC regime at the dc--strength $i_{0}=0.0159$ the
deterministic trajectories, $D=0$,  (blue dotted lines) for large times (upper
panel)  approach  
either a locked solution in one of the wells of the periodic potential
or a running solution that proceeds in the negative direction opposite
to the positive dc--bias. The remaining parameters are the same as in
Fig. 1. The sample trajectories of the Langevin equation (\ref{JJ2})
(red solid lines), resulting for the noise strength $D= 0.001$ and the
same other parameters, stay close to either of the deterministic
trajectories for some time  and then switch to the neighborhood of
another trajectory. The inset of the upper panel displays the locked and the running
deterministic trajectory and one realization of a stochastic
trajectory for five periods of the ac driving.   The corresponding phase portraits for the  deterministic running solution and the stochastic trajectory are shown in lower panels.  
}
\label{fig2}
\end{figure}
\begin{figure}
\includegraphics[width=1\linewidth]{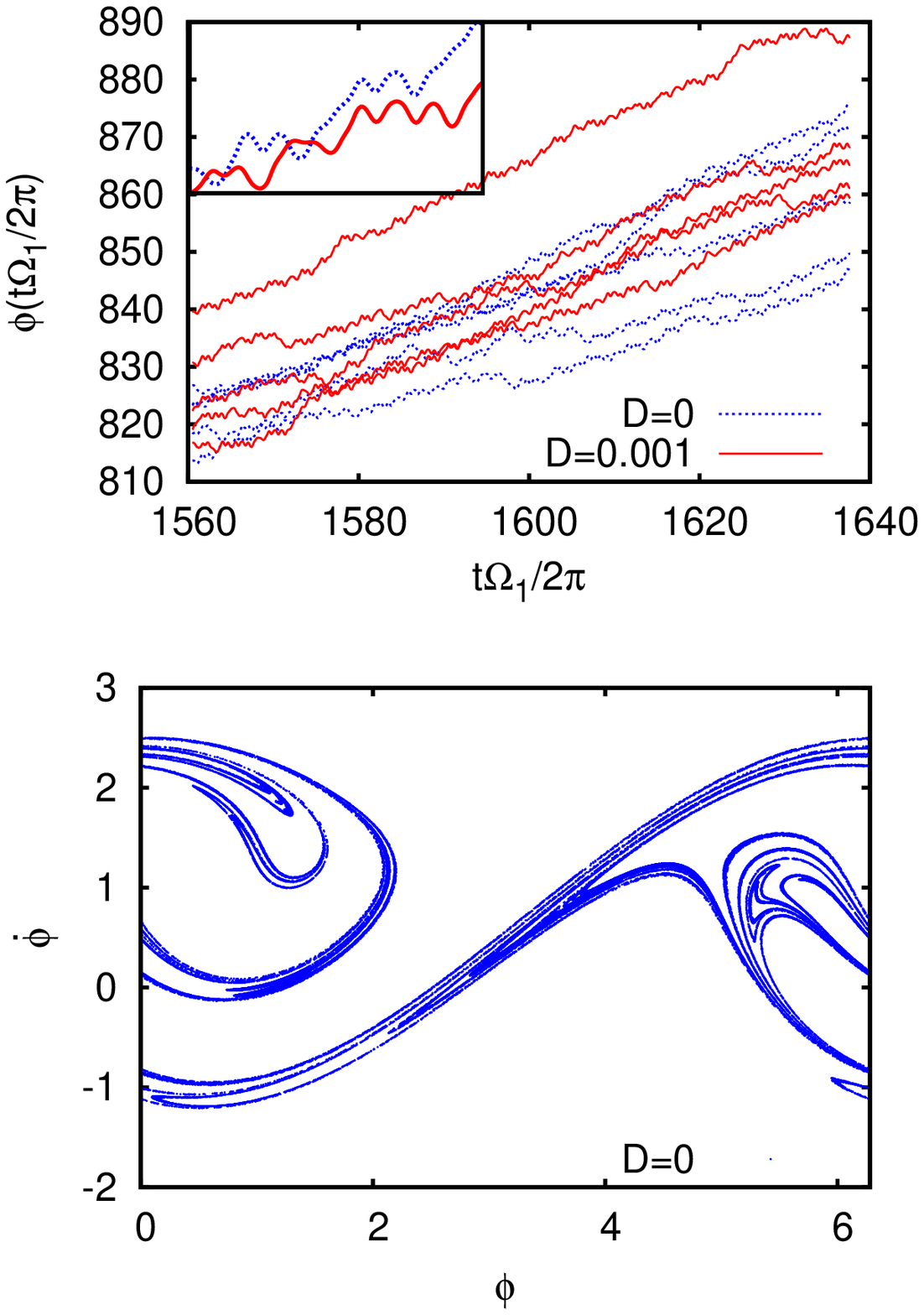} 
\caption{(color online) 
In the Ohmic like regime at $i_0=0.1114$ (other parameters as in the
Figs. 1 and 2) the deterministic motion for $D=0$ (blue dotted line)
is chaotic in the limit of large times. The phase difference $\phi(t)$ 
increases on average, hence leading
to normal conductance, see the upper panel. 
In their qualitative appearance the deterministic trajectories are not
different from the stochastic trajectories (red solid line) 
for the noise strength
$D=0.001$. The Poincar\'e map of the
deterministic system for the same set of parameters is displayed in
the lower panel.  This map represents the 
phase difference modulo $2 \pi$ and the corresponding velocity of a single
trajectory taken at
integer multiples of the ac driving period, see e.g. \cite{Strogatz}.
It reveals the
typical features of a strange attractor.}     
\label{fig3}
\end{figure}
Fig. 1 depicts the main transport characteristics. Panel (a) represents
the dependence of the averaged voltage on the dc--strength.
It displays ANC for small dc--strengths and NDC for larger
dc--values. Fig. 2  of Ref. \cite{kosturPRB} seemingly exhibits a very
similar behavior. The present ac--strength $i_1=0.732$ however is
somewhat larger than 
the one chosen in  Ref. \cite{kosturPRB}. As a consequence, the present set of
parameters 
leads to ANC already for $D=0$.  
The deterministic motion then is governed by two coexisting
attractors, a locked and a non--chaotic running solution, see
Fig. 2.  

In panel (b) of Fig. 1 the standard deviation $\sigma_v$ of the voltage 
is depicted as a function of $i_{0}$. We note
that its dependence on the dc--strength is rather complicated and
without any immediately obvious relation to the averaged voltage of panel (a).  
Upon closer
inspection, one though observes that the voltage fluctuations 
undergo rapid changes yielding a large standard deviation
in the vicinity of zero bias (at small dc--values) and in
the interval where the regime of the NDC appears. In these
regimes, the maxima of the standard deviation are found, cf. the inset in the
upper panel of Fig. 4, where one can reveal oscillations in the
potential well.  Upon a further increase of the dc--strength, the
voltage
fluctuations decrease because the influence of the
periodic potential then becomes weaker.  Since in this limit
oscillations and back--turns of the trajectories become less frequent  the
standard deviation of the voltage decreases. 

\begin{figure}
\includegraphics[width=1\linewidth]{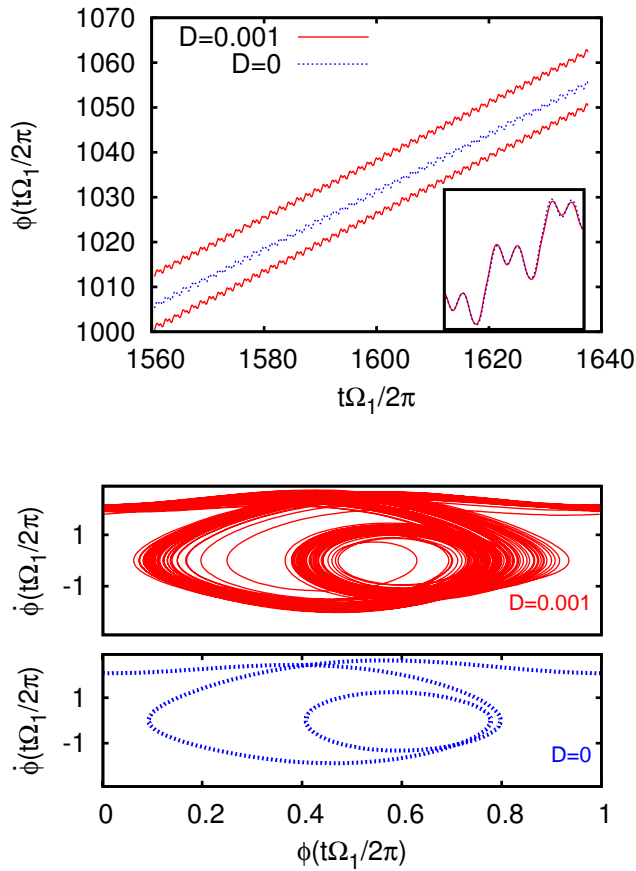} 
\caption{(color online)  
In the regime of NDC at $i_0=0.175$ (all other parameters as in the
previous Figs.) the deterministic motion is represented by 
periodic trajectories (blue dotted lines) running in the positive
direction. 
In this parameter regime the stochastic trajectories at $D=0.001$ (red
solid line) stay much closer 
and for much longer times nearby a deterministic trajectory than in the
cases displayed in Figs 2 and 3 (see the magnification in the inset and 
also compare with the
inset of the upper panel of Fig. 3). This
explains the small value of the phase diffusion coefficient $D_{\phi}$
in this parameter regime, see also panel (c) of Fig. 1. Also in the phase
portrait shown in the lower panel the stochastic
trajectory (red solid curve)  remains most of the time in a close
vicinity of the deterministic trajectory (blue dotted curve).  
}
\label{fig4}
\end{figure}
In panel (c) of Fig. 1 we demonstrate the influence of the dc--strength 
on the P\'eclet number which  is a non-linear function of the
bias $i_0$:   for small, increasing dc--strengths,
it first increases and then 
drops again; the rapid growth of the voltage fluctuations around
$i_{0} \approx 0.175$ is accompanied
by an increasing  P\'eclet number.  In this regime the
trajectories of the noisy dynamics stay almost always very close to the
periodic deterministic attractor, trajectories bundle closely together, 
see the upper panel of Fig. 4.

Finally we study the efficiency of the device. In panel (d) of
Fig. 1, the two efficiency quantifiers, $\eta_E$ and  $\eta_S$ are presented. 
Both vanish if the averaged voltage is zero. 
In the regime outside of ANC, the energy conversion  efficiency
$\eta_E$ is a monotonically increasing  function of the dc--strength.  
In the  regime of the  NDC (in the vicinity of  $i_0=0.175$,
cf. Fig. 1), this efficiency is almost constant with the value
$\eta_{E} = 0.3$.   With a further increase of the dc--strength it increases and saturates 
to the value 1 for large $i_0$.  The Stokes efficiency  $\eta_S$  attains a local minimum in the
vicinity of the dc--strength  $i_0=0.18$ and is 
 always  smaller  than the efficiency of the energy conversion $\eta_E$.  
For large $i_0$, it also approaches the value 1. 
In the regime of ANC (shown in the inset of panel (d) in Fig. 1),  both efficiencies  
are small, of the order $10^{-3}$.  In Refs.
\cite{Suzuki2003},  the rectification efficiency
$\eta_R$ is introduced. Adopting this definition to the system (\ref{JJ2}), we get 
\be
 \eta_R = \frac{-i_0 v + \gamma v^2}{-i_0 v +  \gamma [\la \dot\phi^2 \ra - D]}. 
\ee{etaR}
For the present system,   $\eta_R$  takes both positive and
 negative values. Moreover,   both the corresponding   efficiency of energy conversion $\eta_E$ and   
the Stokes efficiency  $\eta_S$  when evaluated with  the input-denominator as 
given with Eq.  (\ref{etaR}) assume  values larger than unity.  Therefore, these so evaluated three measures  $\eta_R, \eta_E$ and $\eta_S$  are no longer suitable to characterize 'efficiency' 
in the present context with an inertial dynamics determined  from  Eq. (\ref{JJ2}). 

\section{Summary}

Although Josephson junctions have been studied and explored for many
years, still  new   intriguing properties are discovered  in these particular
devices, which also 
have  a great potential to impact novel technologies. They
belong to the most promising candidates for    solid based quantum qubits \cite{qubits}.   
ANC in Josephson junction has recently theoretically been predicted
\cite{machuraPRL} and subsequently
confirmed experimentally \cite{raj}. 
In the present work, we continued the study of  the main
transport characteristics of such  systems. We presented the
voltage-current characteristic which manifests  a regime of ANC and 
two regimes of NDC ( for $i_0 \approx 0.08$ and $i_0 \approx  0.15$). 
These effects may be realized  under various conditions  by a proper 
choice of the system's parameters such as temperature, frequency and ac
amplitude and    dc strength. 
We revealed that the voltage fluctuations characterized by the voltage
standard deviation and the phase difference diffusion coefficient 
 assume a non-monotonic behavior as functions of the
external load. The voltage standard variation exhibits a global maximum in
the second regime of  the NDC, while the diffusion coefficient has a
global minimum in this regime.  Within the ANC regime the 
energetic efficiency is   small while in the regime of the NDC it
takes much larger values.   

\begin{acknowledgments}
\noindent The work supported in part  by the MNiSW Grant
N202 203534 and the Foundation for Polish Science (L. M.).
\end{acknowledgments}

\end{document}